\documentclass[%
 reprint,
 amsmath,amssymb,
 aps,
]{revtex4-2}

\usepackage{graphicx}
\usepackage{graphicx, xurl, hyperref}
\usepackage{color}
\usepackage{dcolumn}
\usepackage{bm}

\begin{document}

\preprint{APS/123-QED}

\title{Liquid-Droplet Coalescence: CNN-based Reconstruction of Flow Fields from Concentration Fields}

\title{Liquid-Droplet Coalescence: CNN-based Reconstruction of Flow Fields from Concentration Fields }
\author{Vasanth Kumar Babu}
	\email[]{vasanthb@iisc.ac.in}
		\affiliation{Centre for Condensed Matter Theory, Department of Physics, Indian Institute of Science, Bangalore, 560012, India. }
    \author{Nadia Bihari Padhan}
     \altaffiliation[Current Address:]{ Institute of Scientific Computing, TU Dresden, 01062 Dresden, Germany}
	 \email[]{nadia_bihari.padhan@tu-dresden.de}
	\affiliation{Centre for Condensed Matter Theory, Department of Physics, Indian Institute of Science, Bangalore, 560012, India. }
	\author{Rahul Pandit}
		\email[]{rahul@iisc.ac.in}
	\affiliation{Centre for Condensed Matter Theory, Department of Physics, Indian Institute of Science, Bangalore, 560012, India. }
 
\date{\today}

\begin{abstract}
Liquid-droplet coalescence and the mergers of liquid lenses are problems of great practical and theoretical interest in fluid dynamics and the statistical mechanics of multi-phase flows. During such mergers, there is an interesting and intricate interplay between the shapes of the interfaces, separating
two phases, and the background flow field. In experiments, it is easier to visualize concentration fields than to obtain the flow field. We demonstrate that two-dimensional (2D) encoder-decoder CNNs, 2D U-Nets, and three-dimensional (3D) U-Nets can be used to obtain flow fields from concentration fields here. To train these networks, we use concentration and flow fields, which we obtain from extensive direct numerical simulations (DNSs) of (a) the coalescence of two circular droplets in the two-component 2D Cahn-Hilliard-Navier-Stokes (CHNS) partial differential equations (PDEs), (b) liquid-lens mergers in the three-component 2D CHNS PDEs, and (c) spherical-droplet coalescence in the two-component 3D CHNS PDEs. We then show that, given test images of concentration fields, our trained models accurately predict the flow fields at both high and low Ohnesorge numbers $Oh$ (a dimensionless ratio of viscous stresses to the inertial and surface-tension forces). Using autoencoders and fully connected neural networks, we also investigate the mapping between the concentration and vorticity fields via low-dimensional latent variables for droplet mergers in the 2D CHNS system. We compare the accuracies of flow-field reconstruction based on the two approaches we employ. Finally, we use data from recent experiments on droplet coalescence to show how our method can be used to obtain the flow field from measurements of the concentration field.
\end{abstract}

\maketitle


\section{\label{sec:Introduction}Introduction}

The importance of artifical intelligence (AI) and data-driven machine learning (ML) is growing exponentially in time as are its applications in investigations of complex phenomena in, e.g., climate-systems science~\cite{knusel2019applying,kashinath2021physics},  fluid flows~\cite{fukami2019super,brunton2020machine,Jaya:PRE2020,pandey2020perspective,jin2020time,guo2016convolutional,kamrava2021physics,santos2020poreflow,liu2021supervised,peng2021data,li2022thermal,xu2023super,li2024synthetic}, collective motion in shoals of fish~\cite{butail2015classification,bhaskar2019analyzing}, and active matter~\cite{cichos2020machine}, to name but a few. Machine-learning models, such as deep neural networks, are increasingly being used to analyse extensive datasets and to increase accuracy in, e.g., classification, prediction, and dimensionality reduction. We show how to use these methods to study the challenging problem of the coalescence of liquid droplets and lenses, which has attracted considerable attention because of its fundamental importance in multi-phase fluid dynamics and statistical mechanics, and its extensive industrial applications. Numerous experimental~\cite{wu2004scaling, aarts2005hydrodynamics, burton2007role, paulsen2011viscous, paulsen2014coalescence,xu2024coalescence} and numerical studies~\cite{eggers1999coalescence, duchemin2003inviscid, gross2013viscous,
khodabocus2018scaling, akella2020universal,padhan2023unveiling} have been devoted to understanding such coalescence. When two droplets or lenses join together, a liquid neck forms between them, and its height $h(t)$ evolves with time $t$ in a manner that depends on the  Ohnesorge number $Oh$, a dimensionless ratio of viscous stresses to the inertial and surface tension forces [$Oh \equiv \nu [\rho/(\sigma R_0)]^{1/2}$ \cite{ohnesorge1936bildung,ohnesorge2019formation,fardin2022spreading}, where $\rho$, $\nu$, $\sigma$ and $R_0$ are, respectively, the density, viscosity, surface tension, and the initial droplet's radius.] In the viscous regime, where $Oh$ is high, $h(t) \sim t$. In contrast, in the inertial regime where $Oh$ is low, $h(t)\sim t^{1/2}$, for spherical droplets, and $h(t)\sim t^{2/3}$, for liquid lenses \cite{burton2007role, paulsen2011viscous, xia2019universality,padhan2023unveiling,scheel2023viscous}. To develop a comprehensive understanding of liquid-droplet or liquid-lens mergers, it is of paramount importance to measure \textit{simultaneously} the concentrations of the immiscible liquids, whose interfaces define droplet and lens boundaries, and the mean flow fields. It is especially challenging to conduct such experiments for several reasons~\cite{kang2004quantitative,minor2007optical,thoroddsen2008high,ortiz2010investigation,castrejon2011dynamics}: (a) to follow the rapid merger of droplets or lenses, high-speed cameras must be used to capture the spatiotemporal evolution of their coalescence; (b) external light sources, commonly used in such measurements, can potentially interfere with and modify the coalescence process; (c) the use of Particle Image Velocimetry (PIV), for the determination of the flow field, is demanding because of the small time scale of coalescence.  

To overcome these challenges, we combine machine-learning (ML) methods with recent advances in direct numerical simulations (DNSs) of the full spatiotemporal evolution of droplet and lens mergers~\cite{palphdthesis,pal2022ephemeral,padhan2023unveiling,scheel2023viscous}, in the incompressible two- or three-component Cahn-Hilliard-Navier-Stokes (CHNS) partial differential equations (PDEs), which we use to model binary- and ternary-fluid mixtures, respectively, in both two and three dimensions (2D and 3D). In particular, we develop encoder-decoder convolution-neural networks (CNNs), which we train with data from our DNSs, to extract the complete flow field from measurements of the concentration fields of the constituents of a multi-phase fluid mixture. To the best of our knowledge, this challenging problem has not been attempted hitherto in a fluid-dynamics context. Given that our DNSs yield fields that match experimental data, our trained CNN should prove to be an invaluable asset for the extraction of flow fields  from concentration fields measured in experiments on liquid-droplet or liquid-lens mergers. We show this explicitly by reconstructing the flow field using illustrative concentration-image data from recent experiments that have been described in Ref.~\cite{xu2024coalescence}.

\section{Results\label{sec:Results}}

\begin{figure*}
    \centering
    \includegraphics[width=\textwidth]{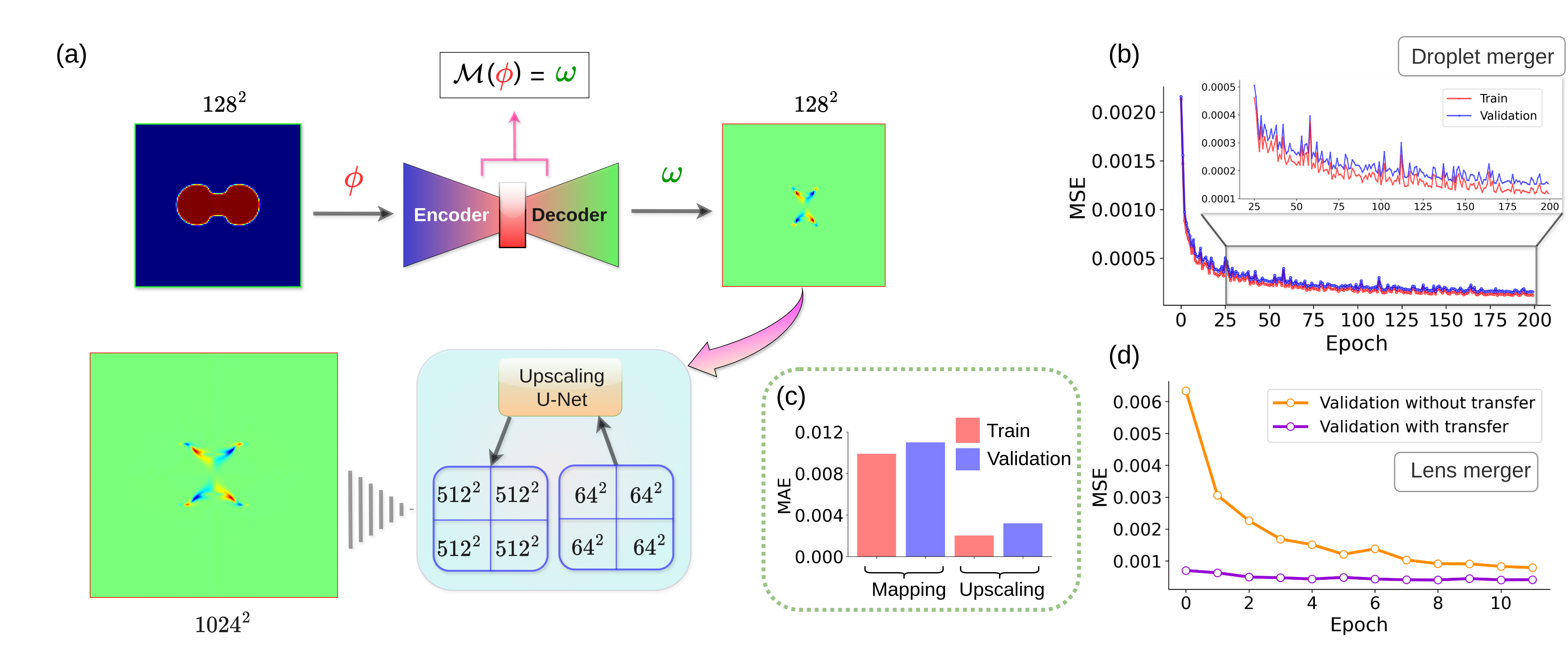}
    \caption{(a) Schematic diagram of our flow-field reconstruction for the merger of two droplets in the 2D CHNS system. We begin with a pseudocolor plot 
of $\phi$, with $128^{2}$ collocation points, at a given time as the input into the encoder-decoder CNN. We train a 2D encoder-decoder CNN to obtain the mapping  ${\mathcal{M}}(\phi)=\omega$. We split the predicted ($128^{2}$) $\omega$ field  into four parts each one of size $64^{2}$, then use these parts, in conjunction with our 2D U-Net, to reconstruct four $\omega$ fields with $512^{2}$  points each, and finally combine these to obtain the full $1024^{2}$ resolution. (b) The plot of MSE, the mean-squared error loss function, for the 2D encoder-decoder CNN, versus the training epochs for the training and the validation data. (c) Mean absolute error [MAE in Eq.~\eqref{eq:MAE}], between the predictions and our DNS data for $\omega$, for both the training and the validation sets, for the mapping and upscaling (see text). (d) Plot of the loss function (MSE), for the 2D encoder-decoder CNN, versus epochs, with (orange curve) and without (purple curve) transfer learning while training for predicting $\omega$ from $\phi$ during a 2D lens merger.
}
    \label{fig:Mapping}
\end{figure*}

Figure~\ref{fig:Mapping} illustrates the essence of our flow-field reconstruction by considering a 2D example of a droplet merger in the 2D CHNS system. We begin with a DNS image of 
the CHNS concentration field $\phi$, at a given time, as the input into the encoder-decoder CNN in Fig.~\ref{fig:Mapping} (a); we use an image with $128^2$ points, which we obtain 
by coarse graining (or resampling) pseudocolor plots of $\phi$ from our DNS with $1024^2$ collocation points. We train a 2D encoder-decoder CNN to predict the corresponding vorticity field $\omega$, on $128^{2}$ collocation points, i.e., we obtain the mapping  ${\mathcal{M}}(\phi)=\omega$ shown in Fig. \ref{fig:Mapping} (a). In Fig.\ref{fig:Mapping} (b), we plot the MSE, the mean-squared error loss function [Eq.~\eqref{eq:loss}], versus the training epochs.
We then train a 2D U-Net to obtain the full-size resolution fields, with $1024^2$ collocation points, from those,
with $128^2$ points, which have been predicted by the 2D encoder-decoder CNN in the previous stage: 
First, we split the predicted ($128^{2}$) $\omega$ field  into four parts of size $64^{2}$. We use these parts, in conjunction with our 2D U-Net, to reconstruct four $\omega$ fields with $512^{2}$  points each [in the Appendix~\ref{app:NNAT}, we describe how the symmetry of the problem can be used to reduce the computations at this stage of reconstruction]. In Fig~\ref{fig:Mapping} (c), we present the mean absolute error [MAE defined in Eqs.~\eqref{eq:MAE}], between the U-Net predictions and our DNS data for $\omega$, for both the training and the validation sets.
We then combine these
$512^{2}$ predictions to obtain the full $1024^{2}$ resolution. 
We give the details of our encoder-decoder CNN, U-net, autoencoder architectures, and training in the Appendix~\ref{app:NNAT}. Henceforth, a caret will indicate the predicted field that we obtain with our U-net; e.g., the predicted vorticity field 
will be denoted by $\hat\omega$.

 \begin{figure*}[htbp]
    \centering
    \includegraphics[width=\textwidth]{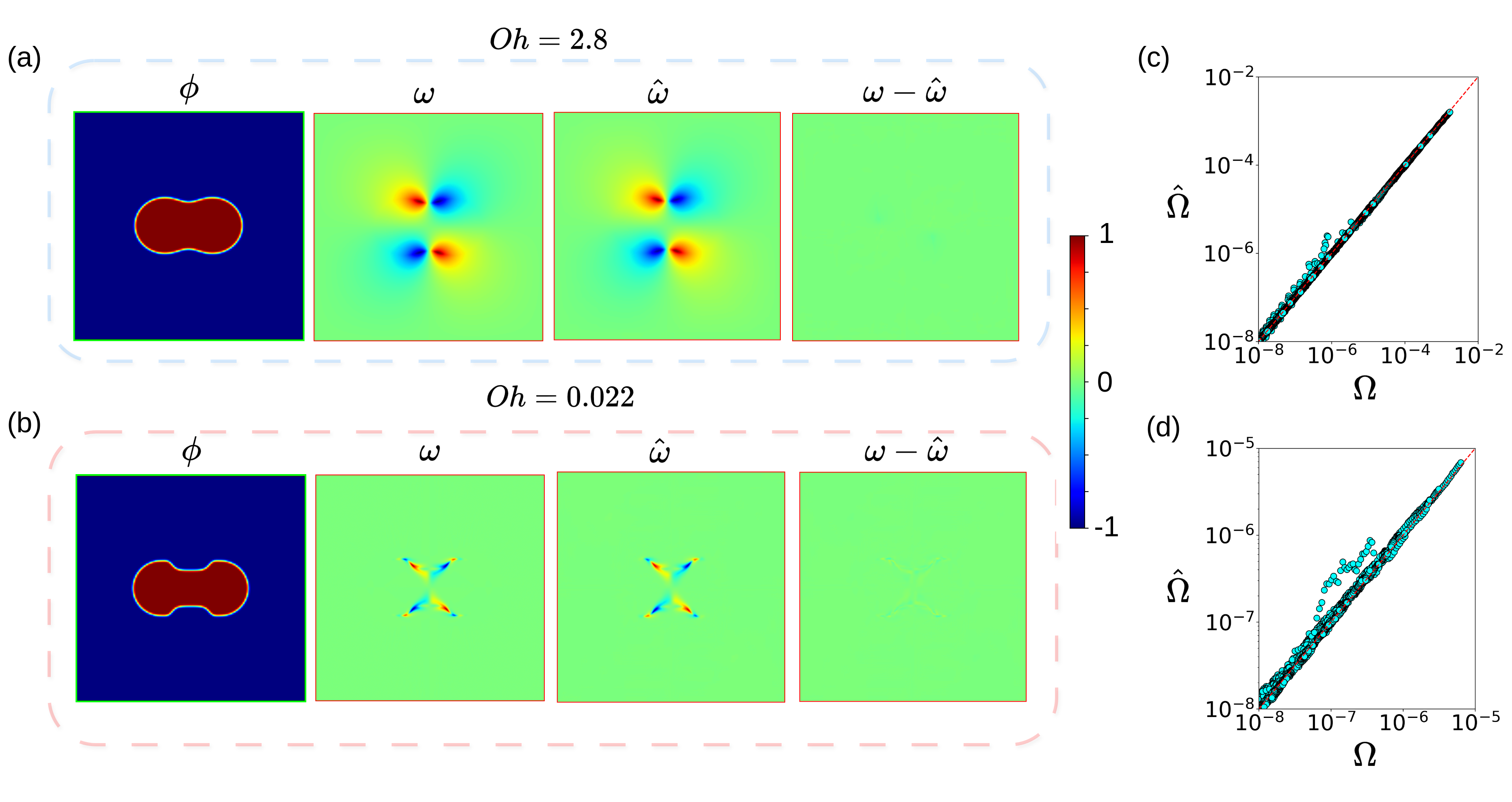}
    \caption{Illustrative comparisons between the $1024^{2}$ vorticity field $\hat\omega$, predicted as in Fig.~\ref{fig:Mapping}, with the ground-truth vorticity field $\omega$, which we obtain from the DNS of a binary-droplet merger in the 2D CHNS system for two representative values of the Ohnesorge number, (a) one high [$Oh=2.8$] and (b) the other low [$Oh=0.022$] from the validation data, and the corresponding plots of the 
difference $\omega - \hat\omega$.   The log-log plots of 
of $\Omega$ versus $\hat\Omega$, for all the values of $k$ and $t$, in (c) and (d),  for  $Oh=2.8$ and $Oh=0.022$, respectively.}
\label{fig:comp_circ}
\end{figure*}

 \begin{figure*}[htbp]
    \centering
    \includegraphics[width=\textwidth]{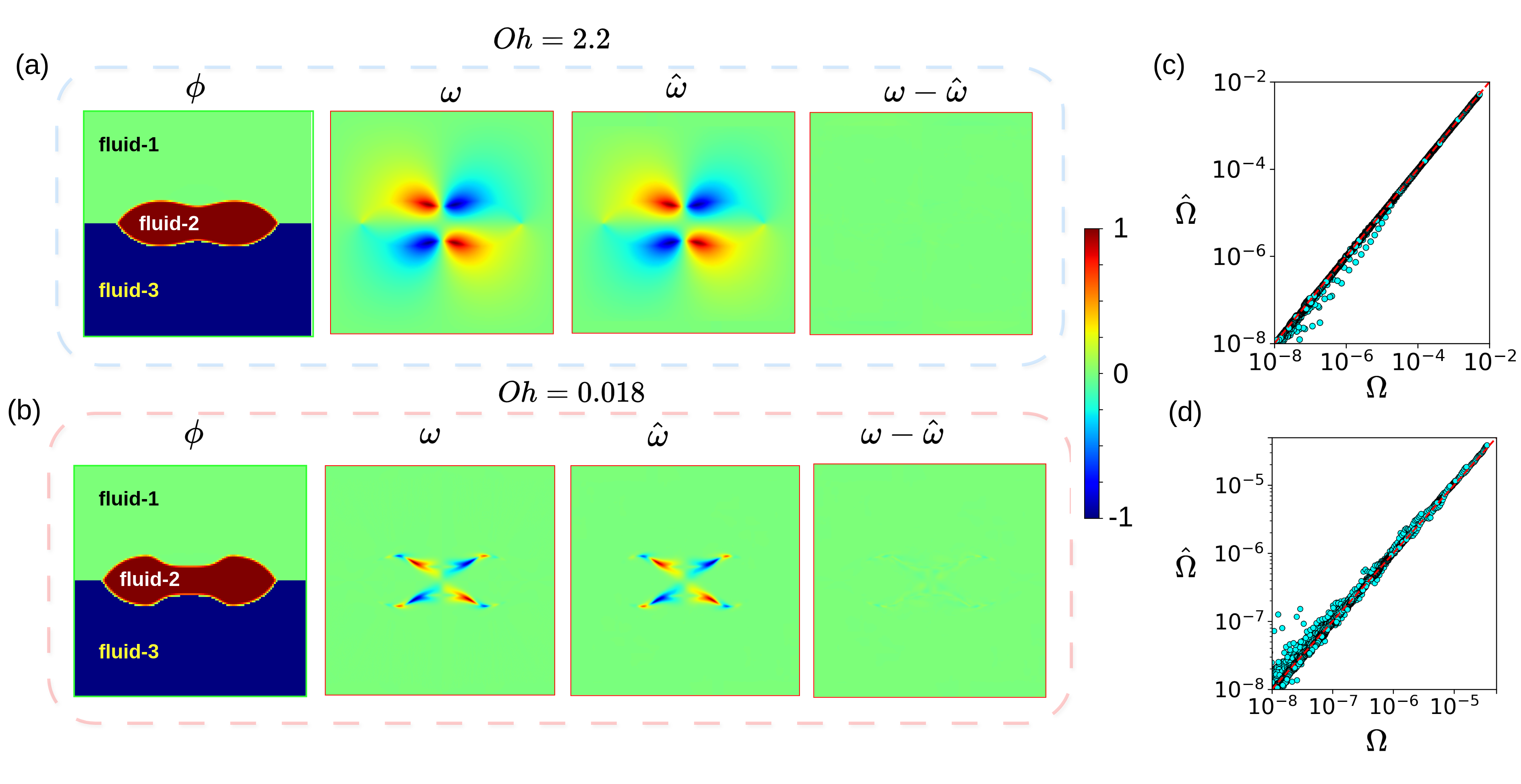}
    \caption{Illustrative comparisons between the $1024^{2}$ predicted vorticity field $\hat\omega$, with the ground-truth vorticity field $\omega$, which we obtain from our DNS of a merger
of two symmetric liquid lenses in the 2D three-component CHNS system for (a) a high value of the
Ohnesorge number [$Oh=2.2$], and (b) a low value [$Oh=0.018$] from the validation data, and plots of the 
difference $\omega - \hat\omega$; (c) and (d) are the  log-log plots of 
of $\Omega$ versus $\hat\Omega$, for all the values of $k$ and $t$.}
    \label{fig:comp_lens}
\end{figure*}

 \begin{figure*}[htbp]
    \centering
    \includegraphics[width=\textwidth]{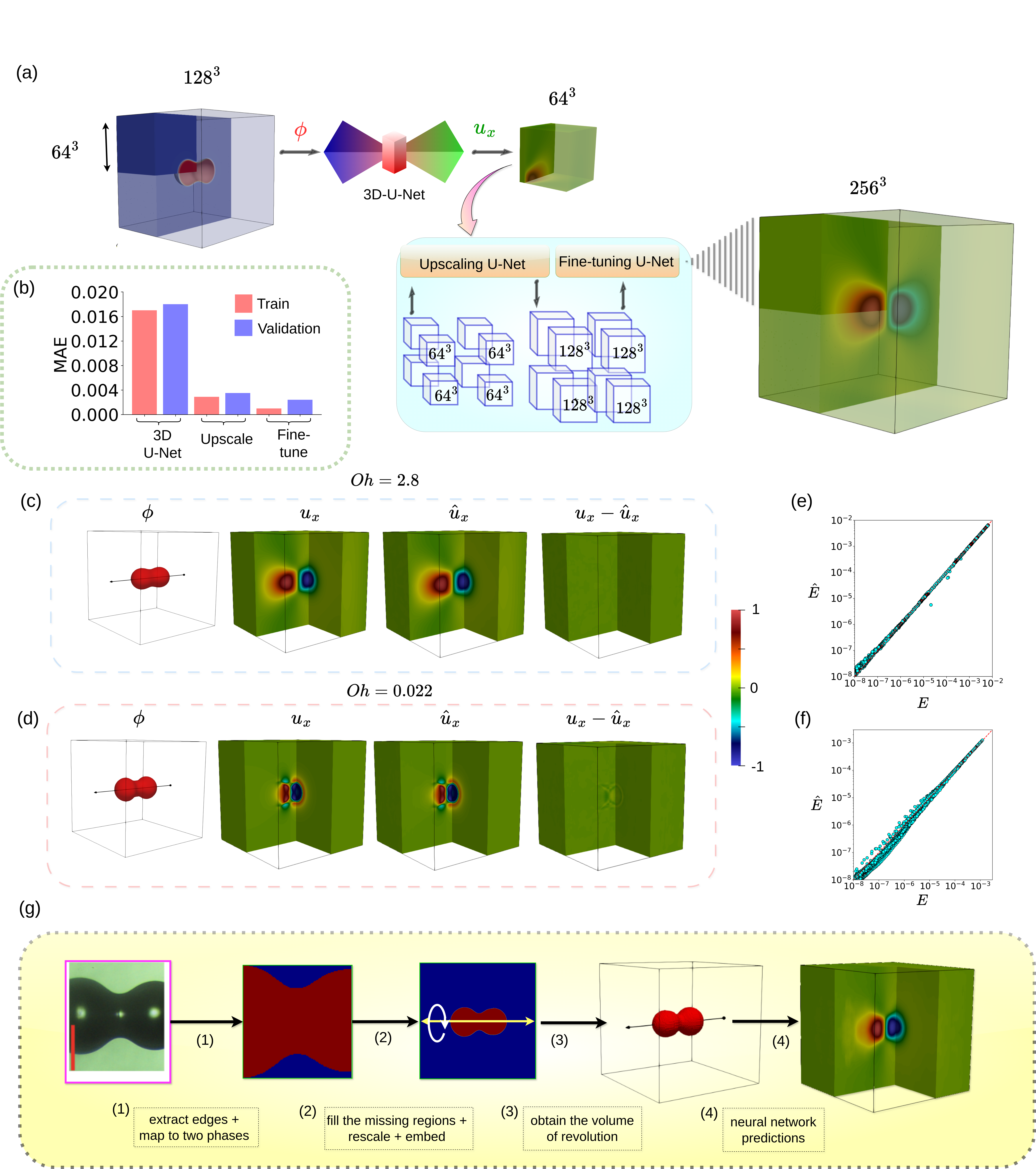}
    \caption{(a) Schematic diagram of the training and reconstruction methods that we employ for the 3D binary droplet merger. (b) Bar chart of the MAE,  between the U-Net predictions and our DNS data for $u_{x}$, for both the training and the validation sets for the 3D-Unet, upscaling 2D U-Net, and the fine tuning (see text). Isosurface plots of $\phi$ and sections through filled contour plots of $u_x$, $\hat{u}_x$, and the 
difference $u_{x} -\hat u_{x}$ for the Ohnesorge numbers $Oh=2.8$ [panel (c)] and $Oh=0.022$ [panel (d)]; (e) and (f) are the log-log plots of $E(k,t)$ [Eq.~\eqref{eq:energy}] that we compute from $u_x$, and $\hat E(k,t)$ that follows from $\hat u_x$. (g) Illustration of the reconstruction for a 2D section of a 3D droplet merger from an experiment [see text and Ref.\cite{xu2024coalescence}].}
    \label{fig:3D_mapping}
\end{figure*}
\paragraph{Circular droplet merger in 2D:} In Fig.~\ref{fig:comp_circ}, we show illustrative comparisons between the $1024^{2}$ vorticity field $\hat\omega$, predicted as in Fig.~\ref{fig:Mapping}, with the ground-truth vorticity field $\omega$, which we obtain from the DNS of a binary-droplet merger in the 2D CHNS system for two representative values of the Ohnesorge number, one high [$Oh=2.8$ in panel (a)] and one low [$Oh=0.022$ in panel (b)], from the validation data. Plots of the 
difference $\omega - \hat\omega$, in Figs.~\ref{fig:comp_circ} (a) and (b), show that it is
negligibly small, so the quality of our prediction for $\hat{\omega}$ is excellent. We also compute the 
shell-averaged fluid enstrophy spectrum $\Omega(k,t)$ [Eq.~\eqref{eq:enstropy}], as a function of the shell wavenumber $k$ and time $t$, with both $\omega$, from our DNS, and $\hat\omega$, from our prediction.
The agreement between these is also excellent, as can be seen from the log-log plots of 
of $\Omega$ versus $\hat\Omega$, for all the values of $k$ and $t$, in Figs.~\ref{fig:comp_circ} (c) and (d),
for  $Oh=2.8$ and $Oh=0.022$, respectively. Had $\Omega$ and $\hat\Omega$ been identical, all points on this plot would
have lain along the red diagonal line. 

\paragraph{Liquid-lens merger in 2D:} We turn now to the merger of two symmetric liquid lenses in 2D. This requires three phases, so we must use the three-component generalization~\cite{boyer2010cahn,padhan2023unveiling} of the CHNS equations [see Models and Methods] to generate our ground-truth fields; in particular, we have the concentrations $c_1,\,c_2,$ and $c_3$ and the vorticity field $\omega$. The concentration fields
must satisfy the constraint $c_1 + c_2 + c_3 = 1$. Lens mergers show up clearly in pseudocolor plots of, e.g., $c_2 - c_1$ [see Fig. 2 in Ref.~\cite{padhan2023unveiling}] that show the three coexisting phases. In the merger of two symmetric lenses, the upper half of the coalescing lenses is a mirror image of the bottom half. To reconstruct $\omega$ from these concentrations, it suffices, \textit{for this symmetric case,} to keep track of only the phase that is inside the merging lenses and the concentration, say $c_2$, of the phase inside the boundary. Given
this simplification, it is natural to use transfer learning and, as a starting point, begin with encoder-decoder and U-Net weights from our previous network, which we have trained above to obtain the voritcity field from the merger of two droplets in a two-component fluid mixture [see Figs.~\ref{fig:Mapping} and \ref{fig:comp_circ}]. In Fig.~\ref{fig:Mapping} (d), the plots of the loss function (MSE) versus epochs, with (orange curve) and without (purple curve) this transfer learning, bring out
clearly the efficacy of such learning.

In Fig.~\ref{fig:comp_lens}, we show illustrative comparisons between the $1024^{2}$ vorticity field $\hat\omega$, predicted as described above, with the ground-truth vorticity field $\omega$, which we obtain from our DNS of a merger
of two symmetric liquid lenses in the 2D three-component CHNS system for a  high
Ohnesorge number [$Oh=2.2$ in panel (a)] and a low one [$Oh=0.018$ in panel (b)],  from the validation data. Plots of the 
difference $\omega - \hat\omega$, in Figs.~\ref{fig:comp_lens} (a) and (b), show that it is
very small; therefore, our prediction $\hat{\omega}$ is excellent. 
The agreement between the spectra $\Omega(k,t)$ and $\hat{\Omega}(k,t)$ is also excellent [see the log-log plots of 
of $\Omega$ versus $\hat\Omega$, for all the values of $k$ and $t$, in Figs.~\ref{fig:comp_lens} (c) and (d)].


\paragraph{Spherical droplet merger in 3D:} We present our results for the merger of two spherical droplets in 
3D, for which we obtain data for the concentration $\phi$ and 
velocity $\bm{u}$ from our DNS of the 3D binary-fluid CHNS system.  The schematic diagram in Fig.~\ref{fig:3D_mapping} (a) gives an overview of the training and reconstruction methods we employ here [see the Appendix~\ref{app:NNAT} for details]:  We first resample the fields $\phi$ and $\bm{u}$ from
$256^{3}$ to $128^{3}$; for specificity in our reconstruction, we concentrate on $u_{x}$, the $x$-component of $\bm{u}$. This $128^{3}$ domain is then divided into eight octants, each of size $64^{3}$. Then, we use these $64^{3}$ fields $\phi$ as the input for a  
3D U-Net, which we train to predict $u_{x}$ of size $64^{3}$  [in the Appendix~\ref{app:NNAT}, we describe how the symmetry of the problem can be used to reduce the computations at this stage]. Given memory constraints, to upsample $u_{x}$ from $64^{3}$ to $128^{3}$, we proceed as follows: With $64^{2}\times4$ slabs as input
along the droplet-merger axis [see the black arrow in the images of $\phi$ in  Figs.~\ref{fig:3D_mapping} (c) and (d)], we then train a 2D-Unet to upsample to slabs of size $128^{2}\times2$;  with a stride of $1$ in the input and a stride of $2$ in the output, this gives the required $64\times2=128$ dimensions along the axis of droplet merger, which can then be combined to obtain the field on $128^{3}$ points. Next we train a 2D U-Net to fine-tune the $128^{2}$ sections of $128^{3}$, with the plane-normals orthogonal to the droplet-merger axis. We follow this by combining these fine-tuned $128^{2}$ sections to obtain fine-tuned $128^{3}$ fields, for each octant. In the last step, we 
combine fields from all the $8$ octants to obtain the final prediction of the field on $256^{3}$ points [for details of the neural network architectures see Tables.~\ref{tab:3d_map},~\ref{tab:3d_ups} and ~\ref{tab:3d_ft} in the Appendix].

In Fig~\ref{fig:3D_mapping} (b), we present a bar chart of the MAE,  between the U-Net predictions and our DNS data for $u_{x}$, for both the training and the validation sets. In Fig.~\ref{fig:3D_mapping}, we show illustrative comparisons between the $256^{3}$ velocity field $\hat u_{x}$, predicted as described above, with the ground-truth velocity field $u_{x}$, which we obtain from our DNS of a droplet merger in our 3D binary-fluid  CHNS system for a  high
Ohnesorge number [$Oh=2.8$ in panel (c)] and a low one [$Oh=0.022$ in panel (d)], from the validation data. Plots of the 
difference $u_{x} -\hat u_{x}$, in Figs.~\ref{fig:3D_mapping} (c) and (d), show that it is very small; therefore our prediction $\hat{u}_{x}$ is very good. 
The agreement between the energy spectrum $E(k,t)$ [Eq.~\eqref{eq:energy}] of $u_x$ and $\hat E(k,t)$ of $\hat u_x$ is also excellent [see the log-log plots of $E$ versus $\hat E$, for all the values of $k$ and $t$, in Figs.~\ref{fig:3D_mapping} (e) and (f)].
 \begin{figure*}
    \centering    \includegraphics[width=\textwidth]{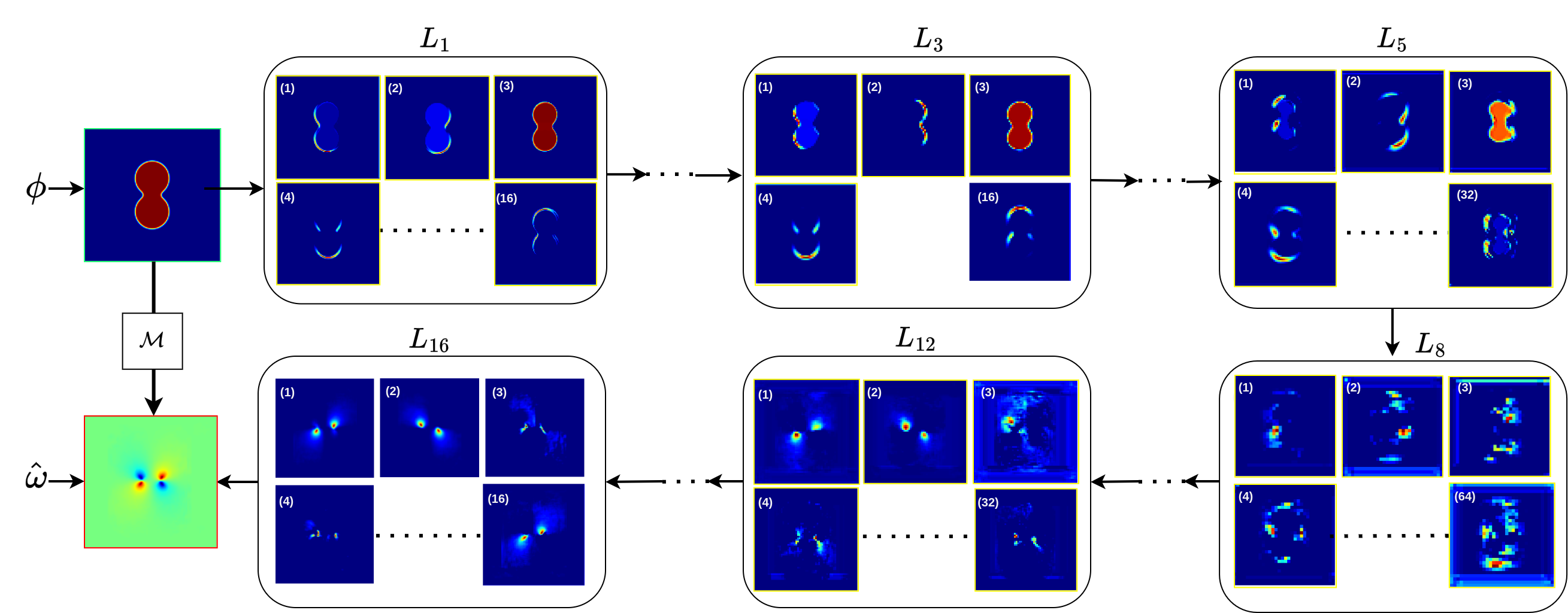}
    \caption{Illustrative pseudocolor plots from hidden layers of our 2D encoder-decoder CNN. In the initial convolutional layer $L_1$, the interface regions (edges) are extracted. The feature maps of $L_3$ still highlight edges, but with reduced spatial dimensions. The $L_5$ layer shows predominantly edge-like structures, but with a slight broadening with traces that are similar to vortices. $L_8$, whose components are the latent variables for this network, shows vortex-like features along with edges. In $L_{12}$, we see low-resolution vortex-like structures. $L_{16}$ has fine vortex-like structures arranged diagonally opposite; these combine to yield the $128^2$ vorticity field $\hat{\omega}$.}
    \label{fig:Hidden_layers}
\end{figure*}
 \begin{figure*}
    \centering    \includegraphics[width=\textwidth]{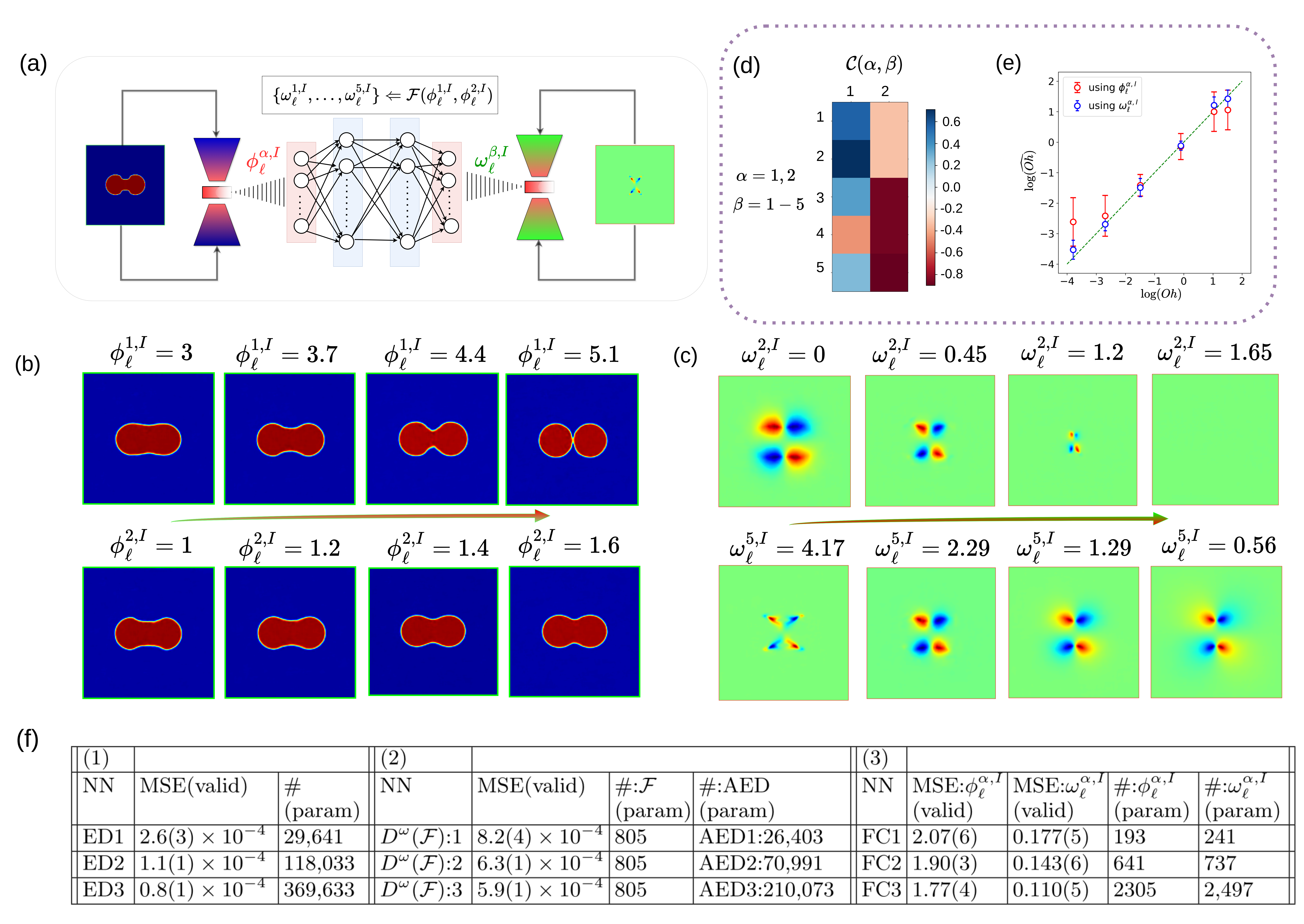}
    \caption{(a) Schematic diagram of voriticity prediction via dimensionality reduction [Eqs.~\eqref{eq:Ephi}-\eqref{eq:Corr}], using autoencoders to obtain the latent variables $\phi^{\alpha,I}_{\ell}$ and $\omega^{\beta,I}_{\ell}$,  the integers $\alpha \in [1,2]$  and $\beta \in [1,2, \ldots, 5]$; the FCNN $\mathcal{F}$ yields the map from $\phi^{\alpha,I}_{\ell}$ to $\omega^{\beta,I}_{\ell}$. (b) Pseudocolor plots of $\phi$, for $128^2$ collocation points, 
with $\phi^{2,I}_{\ell}$ ($\phi^{1,I}_{\ell}$) held fixed at $1.2$ ($3.5$) in the top (bottom) panel; these are used as inputs to $\mathcal{F}$ to obtain $\omega_{\ell}^{\beta,I}$ and the corresponding $128^2$ $\omega$ obtained from the $D^{\omega}$ is given in  the top (bottom) panel of (c). (d)  Pseudocolor plot of the correlation [Eq.~\eqref{eq:Corr}] between the latent variables $\phi^{\alpha,I}_{\ell}$ and $\omega^{\beta,I}_{\ell}$.  (e) Plot of the predicted $\log(\widehat{Oh})$ versus the ground truth $\log({Oh})$. (f) Table (1), for the three encoder-decoders (ED1, ED2, ED3), Table (2), for the three autoencoders (AED1, AED2, AED3), and Table (3), for the three different  FCNNs (FC1, FC2, FC3), showing the mean-square error (MSE),  and the number of parameters (param).} 
 \label{fig:Latent_mapping}
\end{figure*}

We have, so far, used data for $\phi$, from our DNSs of the CHNS PDEs, to obtain the corresponding vorticity or velocity fields.
Can we now use our trained encoder-decoder CNNs to use concentration-image data, from experiments, such as those described in Ref.~\cite{xu2024coalescence}? Yes, indeed, we can, as we illustrate in Fig.~\ref{fig:3D_mapping} (g) for a 2D section of a 3D droplet, which we have obtained from Ref.\cite{xu2024coalescence}. We first obtain the edges from the image, and then set $\phi = 1$ in the region inside it and $\phi = -1$ in the region outside it [this step is indicated by the arrow labelled (1) in Fig.\ref{fig:3D_mapping} (g)]. Our input image is cropped such that it shows only the regions close to the neck of the merging droplets; to obtain the remaining parts of these droplets, we fit circles to the arcs, away from the neck, to obtain a
complete image of the merging droplets [this step is indicated by the arrow labelled (2) in Fig.\ref{fig:3D_mapping} (g)]. We assume that the merging droplets remain axisymmetric over the time scales we consider, so we get the 3D phase field $\phi$ as the volume of revolution obtained by rotating the 2D phase field along the droplet merger axis [this step is indicated by the arrow labelled (3) in Fig.\ref{fig:3D_mapping} (g)],  which we use as the input to our 3D U-net  to obtain the velocity field  [indicated by the arrow labelled (4) in Fig.\ref{fig:3D_mapping} (g)].
\paragraph{Interpreting the CNNs:} To unravel how our encoder-decoder CNN [Table~\ref{tab:2dCNN_map} in the Appendix] predicts $\omega$ from $\phi$, we present illustrative feature maps, for the 2D droplet merger [Fig.~\ref{fig:Hidden_layers}] from various hidden layers of our 2D encoder-decoder CNN. We see an interesting evolution of edges from the initial layer of this CNN to $\omega$ in the final layers. In the initial convolutional layer $L_{1}$ in Fig.~\ref{fig:Hidden_layers}, the interface regions (edges) are extracted; the initial layers capture low-level features such as edges. The $L_3$ layer still highlights edges, but with reduced spatial dimensions. The $L_{5}$ layer shows predominantly edge-like structures, but with a slight broadening with traces that are similar to vortices, i.e., edges act as a structure on which the vorticity is built. The $L_8$, layer, whose components are the latent variables for this network, shows vortex-like features along with edges; their precise roles are difficult to interpret. In $L_{12}$, we observe predominantly low-resolution vortex-like structures. In $L_{16}$ [Fig. ~\ref{fig:Hidden_layers}], the fine vortex-like structures are arranged diagonally opposite; these combine to yield the $128^2$ vorticity field.
\paragraph{Dimensionality reduction:} We now show how to use autoencoders to reduce the dimensionality of the input data for $\phi$ and the output data for  $\omega$, for the mapping part of the problem. Our schema is given in Fig.~\ref{fig:Latent_mapping} (a). For specificity, we illustrate this for the concentration $\phi$ and  vorticity $\omega$ fields on $128^{2}$ collocation points, obtained by resampling  $1024^{2}$-collocation-point data from our DNS of the merger of two circular droplets in 2D CHNS sytem. The encoder part of our autoencoder network performs the following mapping:
\begin{eqnarray}
E^{\phi} &:& \phi^{I}(x,y)\rightarrow\phi^{\alpha,I}_{\ell}\,; \nonumber \\ 
E^{\omega} &:& \omega^{I}(x,y)\rightarrow\omega^{\beta,I}_{\ell}\,;
\label{eq:Ephi}
\end{eqnarray}
here, $(x,y)$ denote the coordinates of the $128^{2}$ collocation points, where $1 \leq I \leq N_D$ labels the data sets (or configurations of $\phi$ and $\omega$)
and $N_D$ is the total number of such data sets, $\phi^{I}(x,y)$ and $\omega^{I}(x,y)$ are the $128^{2}$ concentration and vorticity fields, the integers $\alpha \in [1,2]$  and
$\beta \in [1,2, \ldots, 5]$ label the low-dimensional latent (subscript $\ell$) variables $\phi^{\alpha,I}_{\ell}$ and $\omega^{\beta,I}_{\ell}$ for data set $I$. [See the Appendix~\ref{app:Latent}
for the choice of the range of values for $\alpha$ and $\beta$.]
The decoder part of our autoencoder network performs the inverse of the mapping~\eqref{eq:Ephi}:
\begin{eqnarray}
D^{\phi}:\phi^{\alpha,I}_{\ell}\rightarrow\phi^{I}(x,y)\,; \nonumber \\  
D^{\omega}:\omega^{\beta,I}_{\ell}\rightarrow\omega^{I}(x,y)\,;
\label{eq:Dphi}
\end{eqnarray}
i.e., $D^{\phi}(E^{\phi}(\phi^{I}(x,y)))$$=$$\phi^{I}(x,y)$ and $D^{\omega}(E^{\omega}(\omega^{I}(x,y)))=\omega^{I}(x,y)$, which  we summarize in Fig.~\ref{fig:Latent_mapping} (a). We use $\mathcal{F}$, a fully connected neural network (FCNN), to relate the latent variables
as follows [Fig.~\ref{fig:Latent_mapping} (a)]:
\begin{equation}
\{\omega^{1,I}_{\ell}, \ldots, \omega^{5,I}_{\ell}\} \Leftarrow \mathcal{F}(\phi^{1,I}_{\ell},\phi^{2,I}_{\ell})\,. 
\label{eq:latmap}
\end{equation}
Now we assess the roles played by the latent variables $\phi_{\ell}^{\alpha,I}$ and $\omega_{\ell}^{\beta,I}$. 
In the top panel of Fig.~\ref{fig:Latent_mapping} (b), we fix $\phi^{2,I}_{\ell} = 1.2$ and then increase $\phi^{1,I}_{\ell}$ from $3$ to $5.1$; we then use these values $\phi^{1,I}_{\ell}$ and $\phi^{2,I}_{\ell}$ as inputs to $D^{\phi}$ [Eq.~\eqref{eq:Dphi}], to reconstruct the concentration field $\phi$ with $128^{2}$ collocation points. We see that the changes in $\phi^{1,I}_{\ell}$ lead predominantly to modifications of the width of the neck of the merging droplets. Similarly, in the bottom panel of Fig.~\ref{fig:Latent_mapping} (b),  we fix $\phi^{1,I}_{\ell} = 3.5$ and then increase $\phi^{2,I}_{\ell}$ from $1$ to $1.65$; next we use these values of $\phi^{1,I}_{\ell}$ and $\phi^{2,I}_{\ell}$ as inputs to $D^{\phi}$ [Eq.~\eqref{eq:Dphi}], to reconstruct the concentration field $\phi$ with $128^{2}$ collocation points, from which we surmise that such changes in $\phi^{2,I}_{\ell}$ lead predominantly to the evolution of the shape of the interface especially in the vicinity of the neck of the coalescing droplets. To go from the fields $\phi$ in Fig.~\ref{fig:Latent_mapping} (b) to the corresponding vorticity fields  $\omega$ in Fig.~\ref{fig:Latent_mapping} (c) we obtain the following correlation function between the latent variables:  
\begin{eqnarray}
\mathcal{C}(\alpha,\beta) &=&\frac{\sum_{I=1}^{N_D} [\phi^{\alpha,I}_{\ell} - \bar{\phi}^{\alpha,I}_{\ell}][\omega^{\beta,I}_{\ell} - \bar{\omega}^{\beta,I}_{\ell}]}{N_{D}\sigma_{\alpha}\sigma_{\beta}}\,;\\
\sigma_\alpha &=& \sqrt{\frac{\sum_{I=1}^{N_D} [\phi^{\alpha,I}_{\ell} - \bar{\phi}^{\alpha,I}_{\ell}]^2}{N_D}}\,;\\
\sigma_\beta &=& \sqrt{\frac{\sum_{I=1}^{N_D} [\omega^{\beta,I}_{\ell} - \bar{\omega}^{\beta,I}_{\ell}]^2}{N_D}}\,;
\label{eq:Corr}
\end{eqnarray}
here, the overbars indicate the average over the validation data sets $I$.

With $\phi^{\alpha,I}_{\ell}$ as inputs [Fig.~\ref{fig:Latent_mapping} (b) top panel], we obtain $\omega^{\beta,I}_{\ell}$, as in Eq.\eqref{eq:latmap}, and then reconstruct the corresponding vorticity field $\omega$, with $128^{2}$ collocation points, using $D^{\omega}$ [Eq.~\eqref{eq:Dphi}]. 
From Fig.~\ref{fig:Latent_mapping} (d), we see that $\mathcal{C}(\alpha,\beta)$ is maximal for $\alpha=1$ and $\beta=2$, therefore, we depict 
$\omega^{2,I}_{\ell}$ in the top panel of Fig.~\ref{fig:Latent_mapping} (c), for $\omega^{2,I}_{\ell} = 0,\,0.45,\,1.2$, and $1.65$, which correspond, respectively, to $\phi^{1,I}_{\ell} = 3,\,3.7,\,4.4$, and $5.1$ in the top panel of Fig.~\ref{fig:Latent_mapping} (b).
 
Similarly, we give our reconstructions of the $\omega$ field, with $128^{2}$ collocation points, which we obtain from
$\omega^{5,I}_{\ell}$, in the bottom panel of Fig.~\ref{fig:Latent_mapping} (c), for $\omega^{5,I}_{\ell} = 4.17,\,2.29,\,1.29$, and $0.56$; these correspond, respectively, to $\phi^{2,I}_{\ell} = 1,\,1.2,\,1.4$, and $1.6$ in the bottom panel of Fig.~\ref{fig:Latent_mapping} (b), because
$\mathcal{C}(\alpha,\beta)$ is maximal (in magnitude) for $\alpha=2$ and $\beta=5$  [Fig.~\ref{fig:Latent_mapping} (d)]. 

How does our reconstructions $\hat{\omega}$, which we have presented in Figs.~\ref{fig:Mapping} and ~\ref{fig:comp_circ}, compare with the reconstruction that 
we have obtained via the low-dimensional latent-variables $\phi^{\alpha,I}_{\ell}$ and $\omega^{\beta,I}_{\ell}$?
To address this question, we give in small tables (1) and (2) in Fig.~\ref{fig:Latent_mapping} (f), respectively, a comparison of the MSE, for the predictions of $\omega$, obtained by using these two methods. We carry out these comparisons for three different encoder-decoder architectures [ED1, ED2, and ED3] 
and  three different autoencoder architectures [AED1, AED2, and AED3]; the number of parameters in ED1 [resp ED2][resp ED3] are comparable to those in AED1 [resp AED2][resp AED3].
Our expectation that the reconstruction of $\hat{\omega}$ [Figs.~\ref{fig:Mapping} and ~\ref{fig:comp_circ}] is better than that obtained via the low-dimensional latent-variables $\phi^{\alpha,I}_{\ell}$ and $\omega^{\beta,I}_{\ell}$ is borne out by a comparison of the values of MSE. However,  once we have obtained the latent variables  $\phi^{\alpha,I}_{\ell}$ and $\omega^{\beta,I}_{\ell}$, then the mapping problem via the FCNN $\mathcal{F}$ is greatly simplified, because it requires $\simeq 800-900$ parameters.
In particular, with $\phi^{\alpha,I}_{\ell}$ and $\omega^{\beta,I}_{\ell}$, we can train FCNNs to predict $\log(\widehat{Oh})$, where $Oh$ is the Ohnesorge number
for the validation data sets and the caret denotes the predicted value [see the log-log plots in  Fig.~\ref{fig:Latent_mapping} (e) for the quality of these predictions
with the latent variables $\phi^{\alpha,I}_{\ell}$ and $\omega^{\beta,I}_{\ell}$]; the MSEs for this prediction, for three different FCNNs, are given in small table (3) in  Fig.~\ref{fig:Latent_mapping} (f). 
\vspace{-0.1cm}
\section{\label{sec:conclusions}Discussion and Conclusions}
We have demonstrated how to use AI algorithms to overcome the challenging task of constructing flow fields from concentration fields. We have effectively addressed this challenge using two-dimensional (2D) encoder-decoder CNNs, U-Nets, and three-dimensional (3D) U-Nets. Our direct numerical simulations (DNSs) of the CHNS equations in 2D and 3D have played an essential role in this training, as they have provided the necessary multi-phase fluid-dynamics inputs. Although we have shown this with concentration and flow fields taken from our DNSs of the CHNS equations, our trained deep-learning models can also carry out this flow-field reconstruction in experimental multi-phase flows, as we have shown in Fig.~\ref{fig:3D_mapping} (g), where we have employed our 3D and 2D U-Nets, pre-trained via our DNS study of the merger of two 3D droplets in the CHNS system, to predict the velocity field associated with the merging droplets. This has not been attempted hitherto. Therefore, our method can potentially revolutionise the extraction of flow fields, without direct PIV measurements, from images of the concentration fields in multi-phase fluid systems, as we have illustrated by considering the examples of liquid-droplet coalescence and liquid-lens mergers.

\section{Models and Methods}
\label{sec:methods}
For multi-phase fluid flows we use the framework of the Cahn-Hilliard-Navier-Stokes (CHNS) equations [see, e.g., Refs.~\cite{Jacqmin_1999,Magaletti_2013,badalassi2003computation,Pal_2016,perlekar2017two,pal2022ephemeral,padhan2023unveiling}]; in particular, for the two- and three-phase cases we use the binary-fluid~\cite{Pal_2016,perlekar2017two,pal2022ephemeral} and ternary-fluid~\cite{boyer2010cahn,padhan2023unveiling} CHNS systems. We then carry out a pseudospectral DNS, \`a la Padhan and Pandit~\cite{padhan2023unveiling}, for (a) circular-droplet coalescence in 2D, (b) three-fliuid liquid-lens-merger in 2D, and (c) spherical-droplet coalescence in 3D.
We use these DNSs to obtain $\phi$ and $\omega$ in 2D ($u_x$ in 3D), for cases (a)-(c), which we then use as training and validation data for our machine-learning investigations. 
We carry out DNSs for a wide range of values of the kinematic viscosity $\nu$, so the flows cover both inertial and viscous regimes; the non-dimensional
viscosity is given by the Ohnesorge number $Oh \equiv \nu [\rho/(\sigma R_0)^{1/2}]$ [Table~\ref{tab:param_range}]. 


\subsection{Binary-fluid CHNS model}
\label{subsec:bin_CHNS}

In the binary-fluid CHNS model the scalar-order-parameter $\phi$ distinguishes the two fluids, A and B, with $\phi$ positive (negative) in A-rich (B-rich) regions; the interface between these is diffuse. Hydrodynamics is included by coupling $\phi$ to the velocity field ${\bm u}$ as follows: 
\begin{eqnarray}
    \partial_t \phi + (\bm u \cdot \nabla) \phi&=& M \nabla^2 \mu\,; \label{eq:phi}\\
    \partial_t \omega + (\bm u \cdot \nabla) \omega &=& \nu \nabla^2 \omega -\alpha \omega + [\nabla \times (\mu \nabla c)]\cdot \hat e_z\,;\label{eq:omega}\\
    \nabla \cdot \bm u &=& 0\,; \quad \omega = (\nabla \times \bm u)\cdot \hat e_z\,;\label{eq:incom}
\end{eqnarray}
\begin{eqnarray}
    \mu &=& \left( \frac{\delta {\mathfrak{F}}_{LG}}{\delta \phi}\right)\nonumber \\
    &=& -\frac{3\sigma\epsilon}{2} \nabla^2\phi + \frac{24\sigma}{\epsilon} (\phi - \phi^2)(1 - 2\phi)\,;\label{eq:mu}
\end{eqnarray}
here, the Landau-Ginzburg free-energy functional in the domain $\Omega$ is~\cite{pal2022ephemeral}
\begin{eqnarray}
    {\mathfrak{F}}_{LG}(\phi, \nabla\phi) &=& \int_{\Omega} d\Omega\left[ 12 \frac{\sigma}{\epsilon} F(\phi) +\frac{3}{4}\sigma\epsilon {(\nabla \phi)}^{2}\right]\,,\nonumber\\
    F(\phi) &=& {\phi^{2}(1 - \phi)}^{2}\,,\label{eq:FLGW}
\end{eqnarray}
$\mu$ is the chemical potential, $\alpha$ is the coefficient of friction (often present 
in 2D fluid systems), $\sigma$ is the bare surface tension, and $\epsilon$ is the width of 
the interface.

\subsection{Ternary CHNS model}
\label{sec:ter_CHNS}

In 2D it is convenient to use the vorticity-stream-function formulation for the incompressible Navier-Stokes equation to obtain
\begin{eqnarray}
\partial_t{{\omega}} + ({\bm u}\cdot {\nabla}){{ \omega}} &=& \nu {\nabla}^2
{{\omega}} + {\nabla} \times \left({\sum_{i=1}^{3}\mu_{i} {\nabla} c_{i}}\right)\,,\label{eq:2DCHNSA} \\
 \partial_t{c_{j}} + (\bm{u}.{\nabla})c_{j} &=& \frac{M}{\gamma_{j}}{\nabla}^2 \mu_{j}, \;\; j = 1\; \rm{or}\; 2 \,, 
\label{eq:2DCHNSB}
\end{eqnarray}
where we assume, for simplicity, that all the fluids have the same density $\rho = 1$, kinematic viscosity $\nu$, and mobility $M$, and that $\sigma_{12} = \sigma_{23} = \sigma_{13} \equiv \sigma$. The terms with $\sum_{i=1}^{3}\mu_{i} \nabla c_{i}$ yield the stress on the fluid because of the concentration field $c_i$; and $\mu_i = \left( \frac{\delta \mathfrak{F}}{\delta c_i}\right)$, where
in the domain $\Omega$~\cite{Boyer_2006,padhan2023unveiling},
    \begin{eqnarray}
    {\mathfrak{F}}(\{c_i,\nabla c_i\})&=&\int_{\Omega} d\Omega \left[\frac{12}{\epsilon}F_3(\{c_i\}) + \frac{3\epsilon}{8} \sum_{i=1}^{3}\gamma_i(\nabla c_{i})^2\right]\,,\nonumber\\
    F_3(\{c_i\}) &=& \sum_{i=1}^{3} \gamma_i c_{i}^2(1-c_{i})^2\,,
    \label{eq:fe_ternary}
    \end{eqnarray}
the concentration fields $c_i (i = 1, 2, 3)$  are conserved and satisfy $\sum_{i=1}^{3} c_i = 1$,
and the gradient terms give the surface tensions 
$\sigma_{ij}=\frac{(\gamma_i + \gamma_j)}{2}$ between the phases $i$ and $j$. For the 3D ternary-fluid CHNS equations, see Ref.~\cite{padhan2023unveiling}.

At time $t$, the energy and enstrophy spectra are, respectively,
\begin{eqnarray}
E(k,t) &=& \frac{1}{2} \sum_{k-\frac{1}{2}\le k^{'}\le k+\frac{1}{2}} [\tilde{\boldsymbol{u}}(\boldsymbol{k}',t)]\cdot[\tilde{\boldsymbol{u}}(-\boldsymbol{k}',t)]\,,\label{eq:energy}\\ 
\Omega(k,t) &=& \frac{1}{2} \sum_{k-\frac{1}{2}\le k^{'}\le k+\frac{1}{2}} [\tilde{\boldsymbol{\omega}}(\boldsymbol{k}',t)]\cdot[\tilde{\boldsymbol{\omega}}(-\boldsymbol{k}',t)]\,,
\label{eq:enstropy} 
\end{eqnarray}
where the tildes denote spatial discrete Fourier transforms and $k$
and $k'$ are the moduli of the wave vectors $\boldsymbol{k}$ and $\boldsymbol{k}'$.

\subsection{Data generation and pre-processing}
\label{subsec:datagen}

We use pseudospectral direct numerical simulations (DNSs) to obtain the fields $\phi$ and $\omega$ in 2D (or $u_x$ in 3D) that are required for the training and validations of our NNs. These DNSs use the pseudospectral method~\cite{canuto2012spectral,pal2022ephemeral,padhan2023unveiling} in 2D square or 3D cubical computational domains with periodic boundary conditions; derivatives are evaluated in Fourier space, products of fields are evaluated in real space and are then inverse transformed to Fourier space; aliasing errors, which arises because of the third-order nonlinearities, are removed by using the $1/2-$ dealiasing scheme. In our 2D and 3D DNSs we use $1024^2$ and $256^3$ collocation points, respectively. We cover a range of viscosities (and hence Ohnesorge numbers), 
which span viscous and inertial ranges, to obtain both training and validation data; the values of $Oh$ (black for training sets and red for validation sets) are given in Table~\ref{tab:param_range} in the Appendix. 
In 2D, for each value of $Oh$, we include $100$ configurations  of  $\phi$ and $\omega$, at equally spaced intervals in time, starting from the beginning of droplet or lens coalescence to when the neck height becomes comparable to the diameter of the droplet or the width of the lens. In addition, to develop a robust NN and  to prevent overfitting, we use $100$ more configurations $\phi^{'}$, which we obtain from the initial $100$ by random rotations of $\phi$;
\begin{eqnarray}
\phi'(\mathbf{x'})&=&\phi(\mathbf{x}),\, \quad \mathbf{x'}=\mathcal{R}_d\mathbf{x}\,,
\label{eq:rot}
\end{eqnarray}
where the rotated coordinates $\mathbf{x'}$ are $(x',y')$ in 2D and $(x',y',z')$ in 3D; and the angles in the 2D or 3D rotation matrices $\mathcal{R}_2$
and $\mathcal{R}_3$, respectively, are chosen randomly; thus we include randomness in our training data. In 3D, our machine-learning data include $\phi$ as the input and $u_{x}$ as the output. In 3D, we use $40$ snapshots for each value of $Oh$, starting from separate droplets and up until the neck height becomes comparable to the the diameters of the droplets; then, for each value of $Oh$, we include $10$ randomly rotated configurations.
Each pseudocolor plot, of the input $\phi$  and the output $\omega$, in 2D,  and $u_x$, in 3D, is normalized so that the field intensities lie in the interval  $[-1,1]$.

\subsection{Neural networks and training} 
\label{subsec:NN_train}

Encoder-decoder CNNs~\cite{fukushima1980neocognitron,badrinarayanan2017segnet,krizhevsky2017imagenet,ye2019understanding,minaee2021image} are widely used in
image-to-image mappings~\cite{guo2016convolutional,fukami2019super,jin2020time,santos2020poreflow,kamrava2021physics,liu2021supervised,peng2021data,li2022thermal,xu2023super},
so they are well suited for our task 
of obtaining flow fields ($\omega$ or $u_x$) from the concentration field $\phi$. To capture the nonlinear mapping between $\phi$ and $\omega$ in 2D, we use an encoder-decoder with convolutional layers, where the operations between two such layers can be expressed as
\begin{equation} 
L_{n}^{q}(x,y) = f\left(\sum_{q{'},i,j=1,0,0}^{N_f,h-1,w-1} L^{q{'}}_{n'}(x+i,y+j) F_{i,j,q{'}}^{n,q} + b^{n,q}\right)\,,
\label{eq:conv}
\end{equation} 
where $L^{q{'}}_{n'}(x+i,y+i)$ are the outputs from the order $q{'}$ filter from the previous layer $n{'}$, $L_{n}^{q}(x,y)$
are the outputs from the order-$q$ filter of the current layer ${n}$, and we choose $f$ to be the ReLU activation function~\cite{Nair2010RectifiedLU}. 
For the entries of the filter matrices  $F_{i,j,q{'}}^{n,q}$, of height $h$ and width $w$ and bias $b^{n,q}$, we use Xavier initialization~\cite{glorot2010understanding}; subsequently, these entries are updated during the course of the training~\cite{726791} to optimize the network performance. We use $2\times2$ max-pool filters, alternating with convolutional layers, as we show in the Appendix in Table~\ref{tab:2dCNN_map}. The 
max-pool filters reduces the number of collocation points by a factor of $4$ [$2$ from the height and $2$ from the width] by sliding across the outputs from the previous layer and picking the maximal value, of the concerned field, in the $2\times2$ window in the encoder part of our neural network. Conversely, in the decoder section, upsampling layers perform the inverse of max-pooling and expand the number of collocation points. The full details of the sequence of these operations for our 2D reconstructions are given in Table~\ref{tab:2dCNN_map} in the Appendix. With $\phi$ as the input to the above NN, we update the weights of this NN, during the training, to minimize the following mean-squared error (MSE) loss
\begin{eqnarray}
\text{MSE}&=&\left\langle\frac{1}{N^{2}_{p}}\sum_{x,y=1,1}^{N_p,N_p} \left[{\hat{\omega}(x,y)} - \omega(x,y) \right]^ {2}\right\rangle\nonumber\\
&=& \left\langle\frac{1}{N^{2}_{p}} \sum_{x,y=1,1}^{N_p,N_p} \left[ {\mathcal{M}}(\phi(x,y)) - \omega(x,y) \right]^ {2}\right\rangle
 \,,
\label{eq:loss}
\end{eqnarray}
between the output of the neural network $\hat{\omega}$, and $\omega$ obtained by DNS, to optimize our mapping $\mathcal{M}(\phi)=\hat{\omega}$. In Eq.\eqref{eq:loss}, the summation over the indices $x$ and $y$ is over the $N^{2}_{p}$  collocation points in our square domain; and $\left\langle \cdot \right\rangle$ denotes the average over the training data set. To obtain high-resolution $\omega$ via upscaling [see Fig.~\ref{fig:Mapping}], we use a 2D U-Net~\cite{ronneberger2015u}, a specialised encoder-decoder CNN that is adept at capturing small-scale intricate details. In our U-Net, we introduce additional skip connections [see Ref.~\cite{ronneberger2015u}] by concatenating the feature maps from the encoder network with the layers from the decoder [Table.~\ref{tab:2dUNet_ups}]. To predict $u_{x}$ from $\phi$ in 3D, we use 3D CNNs~\cite{ji20123d,scheinker2020adaptive}, with skip connections, or a 3D U-Net~\cite{cciccek20163d,santos2020poreflow}, which uses $3 \times 3 \times 3$ convolutional filters to capture the spatial features of the fields along all three directions.

To achieve dimensionality reduction for our data set, we use autoencoders~\cite{hinton2006reducing,baldi2012autoencoders,lusch2018deep,ladjal2019pca,brunton2020machine,vlachas2022multiscale,pham2022pca}. The encoder part of our autoencoder network has convolutional and max-pooling layers in the beginning, outputs from which are flattened and passed into dense layers as shown in Table.\ref{tab:AED}. The output from the dense nodes from the final layer of the encoder are the latent variables [the number of these dense nodes is the latent-space dimension], which are then fed into the dense layer of the decoder. A series of convolutional and upsampling operations are then performed to recover the original number of collocation points. We then train to minimize MSE loss between the $\phi$ ( or $\omega$ ) from our DNS given as input to our encoder, and corresponding  output from the decoder.

In some cases [see Fig.~\ref{fig:Mapping}], we use the mean absolute error (MAE) 
\begin{eqnarray}
\text{MAE}&=& \left\langle\frac{1}{N^{2}_{p}}\sum_{x,y=1,1}^{N_p,N_p} | {\hat{\omega_{i}}(x,y)} - \omega(x,y) |  \right\rangle\,.
\label{eq:MAE}
\end{eqnarray}

We have implemented the neural networks with TensorFlow~\cite{abadi2016tensorflow} and have carried out computations on an NVIDIA A100 GPU.

\begin{acknowledgments}  
\hspace{10pt}We thank A. Jayakumar for discusssions, and  the Science and Engineering Research Board
(SERB), and the National Supercomputing Mission, India for support, and the Supercomputer
Education and the Research Centre (IISc) for computational resources.
\end{acknowledgments}

\section*{Data and code availibility}
\hspace{10pt}The data and code utilized in this study can be made available from the authors upon reasonable request.

\appendix

\section{\label{app:NNAT}Neural network architectures and training}

In Table \ref{tab:2dCNN_map}, we give the architecture of our 2D encoder-decoder CNN, used for obtaining ${\mathcal{M}}(\phi)=\omega$ [see Fig.~\ref{fig:Mapping} (a)]. We train these CNNs for $200-250$ epochs with a batch size of $32$ and utilise the Adam optimizer~\cite{kingma2014adam}, with the initial learning rate set to $10^{-3}$.  In Table~\ref{tab:2dCNN_map}, we give the architecture for the 2D encoder-decoder CNN ED2 mentioned in the small table inside Fig.\ref{fig:Latent_mapping} (f). The encoder-decoders ED1 and ED3 are similar to ED2;  the number of filters used in each convolutional layer of ED1 (ED3) is half (double) of those used in ED2; the total number of layers and their ordering and activation functions remain the same as in ED2. We train ED1-ED3 for $200-250$ epochs with batch sizes of $32$ [in the table in Fig.~\ref{fig:Latent_mapping} (f), we present the mean and error estimates of our neural networks for these range of training epochs].

In Table~\ref{tab:2dUNet_ups}, we give the details of the 2D U-Net that we employ for upscaling from $128^{2}$ to $1024^2$
collocation points [see Fig. ~\ref{fig:comp_circ}]. While implementing the upscaling U-Net, we reduce the computation involved in training and predicting by a factor of $2$ as follows: Once the $128^2$ $\omega$ is broken into $4$ $64^{2}$ segments along the $4$ octants, we choose $2$ octants (here the $1$st and $2$nd octant or upper-half plane), and train the CNN to obtain the corresponding $\omega$ on $512^{2}$ collocation points; $\omega$ in the lower-half plane (octants $3$ and $4$) is then obtained via reflection of the field in the upper-half, to obtain the final  $\omega$ on $1024^2$ points.
For this we train the 2D U-Net for $100-150$ epochs, with a batch size of $32$. 

Table~\ref{tab:3d_map} is the 3D counterpart of Table~\ref{tab:2dCNN_map}.
Here, we reduce the computations involved in training and prediction by a factor of $8$ by noting that, once we obtain $u_{x}$ with $64^{3}$ 
collocation points, for, say, octant $1$, then $u_x$ in the remaining $7$ octants follows via sign inversions and reflections. We train the neural network in Table.~\ref{tab:3d_map} with batch sizes of $32$ and for $200-250$ epochs. 

In Table~\ref{tab:3d_ups}, we give the 2D U-Net architecture that we use for upscaling the slabs, with $64^2\times4$ collocation points, into  slabs  with $128^2\times2$ collocation points for $u_x$. This 2D U-Net is trained with a batch size of $256$ for $100-120$ epochs.

In Table~\ref{tab:3d_ft}, we give the 2D U-Net that we use for fine-tuning the $128^{2}$ sections of $128^{3}$ data sets. We train this 2D U-Net with a batch size of $256$ for $100-150$ epochs.

In Table~\ref{tab:AED}, we give the architecture of the autoencoder that we use for the reconstruction of concentration and vorticity fields [see Fig.\ref{fig:Latent_mapping}] on $128^2$ collocation points via the low-dimensional latent variables $\phi^{\alpha,I}_{\ell}$ and $\omega^{\beta,I}_{\ell}$
[Eq.~\eqref{eq:Ephi}]. We train these neural networks for $200-250$ epochs with a batch size of $64$. For the mapping between $\phi^{\alpha,I}_{\ell}$ and  $\omega^{\beta,I}_{\ell}$, we use fully connected neural networks [FCNNs $\mathcal{F}$] with two input nodes and five output nodes and three hidden layers consisting of $32$, $16$, and $8$ nodes. We use the ReLU activation function in all the layers; these networks are trained for $10000-11000$ epochs with a batch size of $64$. In Table~\ref{tab:AED}, we also give the architecture of the autoencoder AED3 mentioned in the small tables in Fig.~\ref{fig:Latent_mapping} (f). In AED1 and AED3, the number of convolutional filters in each  convolutional layer is half and double, respectively, of their counterparts in AED2. These networks are trained for $200-250$ epochs with a batch size of $64$. 

We use three FCNNs, namely, FC1, FC2, and FC3, which we use for predicting $\log(Oh)$ from $\phi^{\alpha,I}_{\ell}$ and $\omega^{\beta,I}_{\ell}$ [see the small table in Fig.~\ref{fig:Latent_mapping} (f)]; these have inputs of size $2$ and $5$ and outputs of size $1$, with two hidden layers with $16\text{\&} 8$, $32\&16$, and $64\&32$ nodes, respectively. We use the ReLU activation in all the layers, except in the final layer where we use a linear-activation function. These FCNNs are trained for $4000-5000$ epochs with a batch size of $64$. In the small tables in Fig.~\ref{fig:Latent_mapping} (f), we present the mean and error estimates of our neural networks for the specified training epochs.\newline\newline
\vspace{-1cm}
\section{\label{app:Latent}Latent space for dimensionality reduction}
In Fig.~\ref{fig:Latent_mapping} of Section~\ref{sec:Results}, we have introduced the latent variables  $\phi^{\alpha,I}_\ell$ and $\omega^{\beta.I}_{\ell}$, with the integers $\alpha = 1\; \rm{or} \; 2$ and $\beta = 1 \ldots 5$. We have chosen these values for these integers for the following reasons. 

For the reconstruction of $\phi$ [Fig.~\ref{fig:Latent_mapping}], if we allow $\alpha$ to go up to $3$, then we find that the MSE errors~\eqref{eq:loss}, for the validation data with $\alpha=1$, $\alpha=1 \; \rm{or} \; 2$, and $\alpha=1 \ldots 3$, are $\simeq8.9(8)\times10^{-3}$, $\simeq3.2(5)\times10^{-3}$, and $\simeq2.5(6)\times10^{-3}$, respectively. When $\alpha=1 \; \rm{or} \; 2$, we observe that, if either $\phi^{1,I}_\ell$ or $\phi^{2,I}_\ell$, in Fig.~\ref{fig:Latent_mapping} (b) is zero for all $I$, then the accuracy of reconstruction is poor [compared to cases in which both $\phi^{1,I}_\ell$ and $\phi^{2,I}_\ell$ are non-zero for all $I$], so we discard these autoencoders. If we allow $\alpha$ to go beyond $3$, then we find that one of 
$\phi^{1,I}_\ell$, $\phi^{2,I}_\ell$, $\phi^{3,I}_\ell$, etc., is zero for all $I$, so no advantage accrues to the reconstruction by increasing the allowed values of $\alpha$ beyond $3$. Our choice of $\alpha=1\; \rm{or} \; 2$ is ideal because it offers a good balance between interpretability and reconstruction accuracy.  

In Table~\ref{tab:vort_recon_MSE},  we show the reconstruction MSE error for $\omega$ using the latent variables $\omega^{\beta,I}_{\ell}$. There is a rapid decrease in the reconstruction error from $\beta=1\; \rm{or} \; 2$ to $\beta=1 \ldots 4$. The reconstruction error does not improve after $\beta=1 \ldots 4$; however, the number of non-zero entries (for all $I$) increases up until $\beta=1 \ldots 9$. Finally we choose $\beta=1 \ldots 5$, because we find that, if we do not include randomly rotated pseudocolor plots in our data set, the maximal number of non-zero entries for $\omega^{\beta,I}_{\ell}$ is $5$.

\begin{table}[htbp] 
    \centering
    \label{tab:param_range}
    \renewcommand{\arraystretch}{1.4}
    \begin{tabular}{|>{\centering\arraybackslash}p{0.8cm} |>{\centering\arraybackslash}p{1.8cm} |>{\centering\arraybackslash}p{3.0cm} |>{\centering\arraybackslash}p{1.6cm}|}
        \hline
        Layer &  Type & Details & Parameters  \\
        \hline
          &  Input & $\phi$ on $128^2$ collocation points & 0 \\
        \hline
        1 &  Conv2D & $3^{2}$ filter + 16 channels + ReLU Activation & 160 \\
        \hline
        2 &  Conv2D & $3^{2}$ filter +  16 channels + ReLU Activation & 2320  \\
        \hline
        3 &  Maxpool2D & $2^{2}$ filter &  0  \\        
        \hline
        4 &  Conv2D & $3^{2}$ filter + 32 channels + ReLU Activation &  4640  \\
        \hline
        5 &  Conv2D & $3^{2}$ filter +  32 channels + ReLU Activation & 9248  \\
        \hline
        6 &  Maxpool2D & $2^{2}$ filter &  0  \\        
        \hline
        7 &  Conv2D & $3^{2}$ filter + 64 channels + ReLU Activation &  18496  \\
        \hline
        8 &  Conv2D & $3^{2}$ filter +  64 channels + ReLU Activation & 36928   \\
        \hline     
        9 &  Upsample2D & $2^{2}$ filter &  0  \\        
        \hline
        10 &  Conv2D & $3^{2}$ filter + 32 channels + ReLU Activation &  18464  \\
        \hline
        11 &  Conv2D & $3^{2}$ filter +  32 channels + ReLU Activation & 9248   \\
        \hline
        12 &  Conv2D & $3^{2}$ filter +  32 channels + ReLU Activation & 9248  \\
        \hline
        13 &  Upsample2D & $2^{2}$ filter &  0  \\
        \hline
        14 &  Conv2D & $3^{2}$ filter + 16 channels + ReLU Activation &  4624   \\
        \hline
        15 &  Conv2D & $3^{2}$ filter +  16 channels + ReLU Activation & 2320   \\
        \hline
        16 &  Conv2D & $3^{2}$ filter +  16 channels + ReLU Activation & 2320   \\
        \hline
        17 &  Conv2D & $1^{2}$ filter +  1 channel + Linear Activation & 17   \\
        \hline               
          &  Output & $\omega$ on $128^2$ collocation points & 0 \\
          \hline
    \end{tabular}
    \caption{The 2D encoder-decoder CNN, which we use to map the $128^{2}$ concentration field $\phi$ to the $128^{2}$ vorticity field $\omega$ for 2D binary-droplet and 2D lens mergers. We give the layer numbers (column 1), their types (column 2), their details (column 3), 
    and the parameters (column 4).}
    \label{tab:2dCNN_map}
\end{table}

\begin{table*}
\begin{minipage}{\columnwidth}
    \centering
    \begin{tabular}{|p{1cm} | p{2.5cm} | p{2.5cm} | p{2cm}|}
        \hline
        Layer &  Type & Details & Parameters  \\
        \hline
         &  Input & $\omega$ on $64^{2}$ collocation points & 0  \\
        \hline
        1 &  Conv2D & $3^{2}$ filter + 32 channels + ReLU Activation & 320  \\
        \hline
        2 &  Conv2D & $3^{2}$ filter +  32 channels + ReLU Activation & 9248 \\
        \hline
        3 &  Maxpool2D & $2^{2}$ filter     &  0  \\        
        \hline
        4 &  Conv2D & $3^{2}$ filter + 48 channels + ReLU Activation &  13872  \\
        \hline
        5 &  Conv2D & $3^{2}$ filter +  48 channels + ReLU Activation & 20784  \\
        \hline
        6 &  Maxpool2D & $2^{2}$ filter   &  0  \\        
        \hline
        7 &  Conv2D & $3^{2}$ filter + 64 channels + ReLU Activation &  27712   \\
        \hline
        8 &  Conv2D & $3^{2}$ filter +  64 channels + ReLU Activation & 36928     \\
        \hline     
        9 &  Upsample2D & $4^{2}$ filter &  0  \\        
        \hline
        10 &  Concatenate & Layer9 + Layer2 &  0  \\
        \hline
        11 &  Conv2D & $3^{2}$ filter +  64 channels + ReLU Activation &  55360  \\
        \hline
        12 &  Conv2D & $3^{2}$ filter +  64 channels + ReLU Activation & 36928   \\
        \hline
        13 &  Upsample2D & $4^{2}$ filter &  0  \\
        \hline
        15 &  Conv2D & $3^{2}$ filter +  48 channels + ReLU Activation & 27696  \\
        \hline
        16 &  Conv2D & $3^{2}$ filter +  48 channels + ReLU Activation & 20784  \\
        \hline
        17 &  Upsample2D & $2^{2}$ filter     &  0  \\
        \hline
        18 &  Conv2D & $3^{2}$ filter + 32 channels + ReLU Activation &  13856   \\
        \hline
        19 &  Conv2D & $3^{2}$ filter +  32 channels + ReLU Activation & 9248   \\
        \hline
        20 &  Conv2D & $3^{2}$ filter +  1 channel + Linear Activation & 289   \\
        \hline  
         &  Output & $\omega$ on $512^{2}$ collocation points & 0   \\
        \hline   
    \end{tabular}
    \caption{The 2D U-Net, which we use to upscale the vorticity field from $128^{2}$ to $1024^{2}$ points for droplet and lens mergers in 2D. We give the layer numbers (column 1), their types (column 2), their details (column 3), 
    and the parameters (column 4).}
    \label{tab:2dUNet_ups}
\end{minipage}\hfill 
\begin{minipage}{\columnwidth}
    \centering
    \begin{tabular}{|p{1cm} | p{2.5cm} | p{2.5cm} | p{2cm}|}
        \hline
        Layer &  Type & Details & Parameters  \\
        \hline
         &  Input & $\phi$ on $64^3$ collocation points& 0  \\
        \hline
        1 &  Conv3D & $3^{3}$ filter + 16 channels + ReLU Activation & 488  \\
        \hline
        2 &  Conv3D & $3^{3}$ filter +  16 channels + ReLU Activation & 6928  \\
        \hline
        3 &  Maxpool3D & $2^{3}$ filter    &  0  \\        
        \hline
        4 &  Conv3D & $3^{3}$ filter + 32 channels + ReLU Activation &  13856  \\
        \hline
        5 &  Conv3D & $3^{3}$ filter +  32 channels + ReLU Activation & 27680  \\
        \hline
        6 &  Maxpool3D & $2^{3}$ filter    &  0  \\        
        \hline
        7 &  Conv3D & $3^{3}$ filter + 64 channels + ReLU Activation &  55360  \\
        \hline
        8 &  Conv3D & $3^{3}$ filter +  64 channels + ReLU Activation & 110656   \\
        \hline     
        9 &  Upsample3D & $2^{3}$ filter    &  0  \\        
        \hline
        10 &  Concatenate & Layer9 + Layer5 &  0  \\
        \hline
        11 &  Conv3D & $3^{3}$ filter + 32 channels + ReLU Activation &  82976  \\
        \hline
        12 &  Conv3D & $3^{3}$ filter +  32 channels + ReLU Activation & 27680   \\
        \hline
        13 &  Conv3D & $3^{3}$ filter +  32 channels + ReLU Activation & 27680   \\
        \hline
        14 &  Upsample3D & $2^{3}$ filter     &  0  \\
        \hline
        15 &  Concatenate & Layer14 + Layer2 & 0  \\
        \hline
        16 &  Conv3D & $3^{3}$ filter + 16 channels + ReLU Activation &  20752   \\
        \hline
        17 &  Conv3D & $3^{3}$ filter +  16 channels + ReLU Activation & 6928   \\
        \hline
        18 &  Conv3D & $3^{3}$ filter +  16 channels + ReLU Activation & 6928   \\
        \hline
        19 &  Conv3D & $1^{3}$ filter +  1 channel + Linear Activation & 17   \\
        \hline  
         & Output  & $u_x$ on $64^3$ collocation points & 0   \\
        \hline
    \end{tabular}
    \caption{The 3D U-Net,  which  we use to map the $128^{3}$ concentration field $\phi$  to the $128^{3}$ velocity field $u_x$ for a droplet merger
    in 3D. We give the layer numbers (column 1), their types (column 2), their details (column 3), 
    and the parameters (column 4).}
    \label{tab:3d_map}
\end{minipage}\hfill 
\end{table*}

\begin{table*}
\begin{minipage}{\columnwidth}
    \centering
    \begin{tabular}{|p{1cm} | p{2.5cm} | p{2.5cm} | p{2cm}|}
        \hline
        Layer &  Type & Details & Parameters  \\
        \hline
         &  Input & $u_x$ on $64^{2}\times4$ collocation points & 0  \\
        \hline
        1 &  Conv2D & $3^{2}$ filter + 32 channels + ReLU Activation & 1184  \\
        \hline
        2 &  Conv2D & $3^{2}$ filter +  32 channels + ReLU Activation & 9248 \\
        \hline
        3 &  Maxpool2D & $2^{2}$ filter &  0  \\        
        \hline
        4 &  Conv2D & $3^{2}$ filter + 48 channels + ReLU Activation &  13872  \\
        \hline
        5 &  Conv2D & $3^{2}$ filter +  48 channels + ReLU Activation & 20784  \\
        \hline
        6 &  Maxpool2D & $2^{2}$ filter &  0  \\        
        \hline
        7 &  Conv2D & $3^{2}$ filter + 64 channels + ReLU Activation &  27712   \\
        \hline
        8 &  Conv2D & $3^{2}$ filter +  64 channels + ReLU Activation & 36928     \\
        \hline     
        9 &  Upsample2D & $2^{2}$ filter &  0  \\        
        \hline
        10 &  Concatenate & Layer9 + Layer5 &  0  \\
        \hline
        11 &  Conv2D & $3^{2}$ filter + 64 channels + ReLU Activation &  64576  \\
        \hline
        12 &  Conv2D & $3^{2}$ filter +  64 channels + ReLU Activation & 36928   \\
        \hline
        13 &  Conv2D & $3^{2}$ filter +  64 channels + ReLU Activation & 36928   \\
        \hline
        14 &  Upsample2D & $2^{2}$ filter &  0  \\
        \hline
        15 &  Concatenate & Layer14 + Layer2 & 0  \\
        \hline
        16 &  Conv2D & $3^{2}$ filter + 48 channels + ReLU Activation &  41520   \\
        \hline
        17 &  Conv2D & $3^{2}$ filter +  48 channels + ReLU Activation & 20784   \\
        \hline
        18 &  Conv2D & $3^{2}$ filter +  48 channels + ReLU Activation & 20784  \\
        \hline
        19 &  Upsample2D & $2^{2}$ filter &  0  \\
        \hline
        21 &  Conv2D & $3^{2}$ filter + 32 channels + ReLU Activation &  13856   \\
        \hline
        22 &  Conv2D & $3^{2}$ filter +  32 channels + ReLU Activation & 9248   \\
        \hline
        23 &  Conv2D & $3^{2}$ filter +  32 channels + ReLU Activation & 9248   \\
        \hline
        24 &  Conv2D & $1^{2}$ filter +  2 channels + Linear Activation & 66 \\
        \hline    
         &  Output & $u_{x}$ on $128^{2}\times2$ collocation points & 0   \\
        \hline 
    \end{tabular}
    \caption{The 2D U-Net, which we use to upscale the slabs of $u_x$,  from $64^{2}\times4$ to $128^{2}\times2$, in 3D. We give the layer numbers (column 1), their types (column 2), their details (column 3), 
    and the parameters (column 4)} 
    \label{tab:3d_ups}
\end{minipage}\hfill 
\begin{minipage}{\columnwidth}
    \centering
    \begin{tabular}{|p{1cm} | p{2.5cm} | p{2.5cm} | p{2cm}|}
        \hline
        Layer &  Type & Details & Parameters  \\
        \hline
         &  Input & $u_x$ on $128^2$ collocation points& 0  \\
        \hline
        1 &  Conv2D & $3^{2}$ filter + 32 channels + ReLU Activation & 320  \\
        \hline
        2 &  Conv2D & $3^{2}$ filter +  32 channels + ReLU Activation & 9248 \\
        \hline
        3 &  Maxpool2D & $2^{2}$ filter    &  0  \\        
        \hline
        4 &  Conv2D & $3^{2}$ filter + 48 channels + ReLU Activation &  13872  \\
        \hline
        5 &  Conv2D & $3^{2}$ filter +  48 channels + ReLU Activation & 20784  \\
        \hline
        6 &  Maxpool2D & $2^{2}$ filter     &  0  \\        
        \hline
        7 &  Conv2D & $3^{2}$ filter + 64 channels + ReLU Activation &  27712   \\
        \hline
        8 &  Conv2D & $3^{2}$ filter +  64 channels + ReLU Activation & 36928     \\
        \hline     
        9 &  Upsample2D & $2^{2}$ filter     &  0  \\        
        \hline
        10 &  Concatenate & Layer9 + Layer5 &  0  \\
        \hline
        11 &  Conv2D & $3^{2}$ filter + 48 channels + ReLU Activation &  48432  \\
        \hline
        12 &  Conv2D & $3^{2}$ filter + 48 channels + ReLU Activation & 20784   \\
        \hline
        13 &  Conv2D & $3^{2}$ filter + 48 channels + ReLU Activation & 20784   \\
        \hline
        14 &  Upsample2D & $2^{2}$ filter   &  0  \\
        \hline
        15 &  Concatenate & Layer14 + Layer2 & 0  \\
        \hline
        16 &  Conv2D & $3^{2}$ filter + 32 channels + ReLU Activation &  23072   \\
        \hline
        17 &  Conv2D & $3^{2}$ filter + 32 channels + ReLU Activation & 9248   \\
        \hline
        18 &  Conv2D & $3^{2}$ filter +  32 channels + ReLU Activation & 9248  \\
        \hline
        19 &  Conv2D & $3^{2}$ filter +  1 channel + Linear Activation & 289   \\
        \hline        
          &  Output & $u_x$ on $128^2$ collocation points& 0  \\
        \hline
    \end{tabular}
    \caption{The 2D U-Net, which we  use to fine tune the  $128^2$ sections of $128^3$ predictions of $u_x$ in 3D. We give the layer numbers (column 1), their types
(column 2), their details (column 3), and the
parameters (column 4).}
    \label{tab:3d_ft}
\end{minipage}\hfill
\end{table*}

\begin{table}[htbp] 
    \centering
    \renewcommand{\arraystretch}{1.4}
    \label{tab:param_range}
    \begin{tabular}{|>{\centering\arraybackslash}p{0.8cm} |>{\centering\arraybackslash}p{1.8cm} |>{\centering\arraybackslash}p{4.0cm} |>{\centering\arraybackslash}p{1.6cm}|}
        \hline
        Layer &  Type & Details & Parameters  \\
        \hline
        &  Input & $\phi$ or $\omega$ on $128^{2}$ collocation points & 0  \\
        \hline 
        1 &  Conv2D & $3^{2}$ filter + 64 channels + Linear Activation& 640  \\
        \hline
        2 &  Conv2D & $3^{2}$ filter +  64 channels + ReLU Activation& 36928 \\
        \hline
        3 &  Maxpool2D & $2^{2}$ filter&  0  \\        
        \hline
        4 &  Conv2D & $3^{2}$ filter + 32 channels + ReLU Activation &  18464  \\
        \hline
        5 &  Conv2D & $3^{2}$ filter +  32 channels + ReLU Activation & 9248  \\
        \hline
        6 &  Maxpool2D & $2^{2}$ filter    &  0  \\ 
        \hline
        7 &  Conv2D & $3^{2}$ filter + 16 channels + ReLU Activation &  4624   \\
        \hline
        8 &  Conv2D & $3^{2}$ filter +  16 channels + ReLU Activation & 2320     \\
        \hline     
        9 &  Maxpool 2D & $4^{2}$ filter    &  0  \\        
        \hline
        10 &  Flatten &  - - -    &  0  \\        
        \hline
        11 &  Dense layer &  64 nodes + ReLU activation    &    65600\\        
        \hline
       12 &  Dense layer: Latent space &  2  or 5\, nodes [Latent variables] + ReLU activation    &  130 or 325  \\        
        \hline
        13 &  Dense layer &  256 nodes + ReLU activation &  768 or  1536 \\
        \hline
14 &  Reshaping & $16\times16$ &  0  \\
        \hline
        15 &  Conv2D & $3^{2}$ filter + 16 channels + ReLU Activation &  160 \\
        \hline
        16 &  Conv2D & $3^{2}$ filter +  16 channels + ReLU Activation & 2320   \\
        \hline
        17 &  Upsample2D &    $4^{2}$ filter &  0  \\
        \hline
        18 &  Conv2D & $3^{2}$ filter + 32 channels + ReLU Activation &  4640   \\
        \hline
        19 &  Conv2D & $3^{2}$ filter +  32 channels + ReLU Activation & 9248   \\
        \hline
        20 &  Upsample2D & $2^{2}$ filter   &  0  \\
        \hline
        21 &  Conv2D & $3^{2}$ filter + 64 channels + ReLU Activation &  18496  \\
        \hline
        22 &  Conv2D & $3^{2}$ filter +  64 channels + ReLU Activation & 36928   \\
        \hline
        23 &  Conv2D & $3^{2}$ filter +  1 channel + Linear Activation & 577   \\
        \hline   
         &  Output & $\phi$ or $\omega$ on $128^{2}$ collocation points & 0   \\
        \hline 
    \end{tabular}
   \caption{The autoencoder network, which we use to find the low-dimensional latent variables for the $128^{2}$ concentration $\phi$ and vorticity 
   $\omega$ fields.  We give the layer numbers (column 1), their types
(column 2), their details (column 3), and the
parameters (column 4).}
\label{tab:AED}
\end{table}
\section{\label{app:TVDP}Training and validation data parameters}
 In Table. \ref{tab:param_range}, we give the  values of the  Ohnesorge number $Oh=\nu[\rho/(\sigma R_0)]^{1/2}$ used in our training data set (entries in black) and validation data set (red entries); circles indicate 2D binary droplet coalescence, lenses indicate 2D lens mergers, and spheres 3D droplet coalescence.

\begin{table}[htbp] 
    \centering
    \renewcommand{\arraystretch}{1.5}
    \begin{tabular}{|p{1.6cm}|p{0.9cm} |p{0.9cm}| p{0.9cm}| p{0.9cm}| p{0.9cm}| p{0.9cm}|}
        \hline
        $\beta$ &  $1-2$& $1-3$ & $1-4$& $1-5$ & $1-6$& $1-7$\\
        \hline
    \text{MSE}$\times10^{-4}$& $4.8(4)$& $2.2(3)$&  $0.8(2)$& $0.8(3)$&  $0.8(2)$& $0.8(3)$\\
        \hline
    \end{tabular}
    \caption{The mean square error MSE [row 2] for the validation data for the reconstruction of the vorticity field $\omega$ using the autoencoder
    [Table~\ref{tab:AED} in the Appendix] for different latent-space dimensions [row 1].}
    \label{tab:vort_recon_MSE}
\end{table}

\begin{table}[htbp] 
    \centering
    \renewcommand{\arraystretch}{1.2}
    \begin{tabular}{|>{\centering\arraybackslash}p{1cm} |>{\centering\arraybackslash}p{1.8cm} |>{\centering\arraybackslash}p{1.8cm} |>{\centering\arraybackslash}p{2cm}|}
        \hline
         No& circle 2D:$Oh$  & lens 2D:$Oh$ & sphere 3D:$Oh$ \\
        \hline
        1 &  \textcolor{red}{$0.022$}& \textcolor{red}{$0.018$}& \textcolor{red}{$0.022$}\\
        \hline
        2 & $0.034$& $0.026$&  $0.034$\\
        \hline 
        3 & $0.045$& $0.036$&  $0.045$\\
        \hline
        4 & $0.056$& $0.045$ &  $0.056$ \\
        \hline
        5 & \textcolor{red}{$0.068$}& \textcolor{red}{$0.054$}& \textcolor{red}{$0.068$}\\
        \hline
        6 & $0.079$& $0.062$ &  $0.079$\\
        \hline 
        7 & $0.09$& $0.071$&  $0.09$\\
        \hline
        8 & $0.1$& $0.08$ &  $0.1$ \\
        \hline
        9 & $0.11$& $0.09$&  $0.11$\\
        \hline
        10 & \textcolor{red}{$0.22$}& \textcolor{red}{$0.18$} & \textcolor{red}{$0.22$}\\
        \hline 
        11 & $0.34$& $0.27$&  $0.34$\\
        \hline
        12 & $0.45$& $0.36$ &  $0.45$ \\
        \hline 
        13 & $0.56$& $0.45$&  $0.56$\\
        \hline
        14 & $0.68$& $0.53$ &  $0.68$\\
        \hline 
        15 & $0.79$& $0.62$&  $0.79$\\
        \hline
        16 & \textcolor{red}{$0.9$}& \textcolor{red}{$0.7$} & \textcolor{red}{$0.9$}\\
        \hline 
        17 & $1$ & $0.8$&  $1$\\
        \hline
        18 & $1.1$& $0.9$ &  $1.1$\\
        \hline 
        19 & $1.7$& $1.3$&  $1.7$\\
        \hline
        20 & $2.3$& $1.8$ &  $2.3$ \\
        \hline     
        21 & \textcolor{red}{$2.8$}& \textcolor{red}{$2.2$}&  \textcolor{red}{$2.8$}\\
        \hline
        22 & $3.4$& $2.7$ &  $3.4$\\
        \hline 
        23 & $3.9$& $3.1$&  $3.9$\\
        \hline
        24 & \textcolor{red}{$4.5$}& \textcolor{red}{$3.56$}&  \textcolor{red}{$4.5$}\\
        \hline
        25 & $5.1$& $4$ &  $5.1$ \\
        \hline 
        26 & $5.7$& $4.5$ &  $5.7$ \\
        \hline
    \end{tabular}
    \caption{The values of the  Ohnesorge number $Oh=\nu[\rho/(\sigma R_0)]^{1/2}$ used in our training data set (entries in black) and validation data 
    set (red entries); circles indicate 2D droplet coalescence, lenses indicate 2D lens mergers, and spheres 3D droplet coalescence.}
    \label{tab:param_range}
\end{table}
\clearpage
\bibliography{apssamp}

\end{document}